\input psfig.sty
\documentclass{aa}
\begin{document}

   \thesaurus{08.16.7}  
   
   \title{Optical observations of PSR1706$-$44 with the Test Camera of
the VLT-UT1.\thanks{Based on observations collected at the European Southern Observatory,  
Paranal, Chile (VLT-UT1 Science Verification Program)}}

\author{R.P. Mignani\inst{1}, P. A. Caraveo\inst{2,3} and
G.F. Bignami\inst{4,5}} 

   \offprints{R.P. Mignani}

   \institute{STECF-ESO, Karl Schwarzschild Str.2  D85740-Garching, Germany \\
         email:rmignani@eso.org
         \and
         Istituto di Fisica Cosmica del CNR, Via Bassini 15,
         20133-Milan, Italy  \\
	 email:pat@ifctr.mi.cnr.it
         \and 
	 Istituto Astronomico, Via Lancisi 29, 00161-Rome, Italy 
	 \and
  	 Agenzia Spaziale Italiana, Via di Villa Patrizi 13,
         00161-Rome, Italy \\
         email:bignami@asi.it
          \and
         Universit\'a di Pavia, Via Bassi 6,  27100-Pavia, Italy 
         }

 \date{Received  13 Oct 1998; accepted 9 Nov 1998}

\titlerunning{VLT observations of PSR1706$-$44.}
\authorrunning{R.P. Mignani et al.}
\maketitle

\begin{abstract}
We   report on  the optical  observations   of the $\gamma$-ray pulsar
PSR1706$-$44 performed with the Test Camera of  the VLT-UT1 during the
telescope Science  Verification programme.  With a  limiting magnitude
of $V 
\sim 27.5$, these are the deepest optical observations ever performed
for this pulsar,  for which, however, no  likely counterpart  has been
detected.  The non-detection  of PSR1706$-$44 sets  an upper limit  on
its  optical luminosity varying  from  $\simeq  2~10^{28}$ to  $\simeq
5~10^{29}$ erg s$^{-1}$, depending on the exact amount of interstellar
absorption. 
   
   \end{abstract}

%

\section{Introduction}

PSR1706$-$44 is a  young   ($\simeq$ 17\,000 yrs)   pulsar ($P$=102~ms),
discovered  during  a 20cm  radio survey  of  the southern  hemisphere
(Johnston et al. 1992).  The pulsar was originally associated with the
supernova remnant  MSH15-52 (McAdam, Osborne  and Parkinson, 1993) but
pulsar scintillation measurements  (Nicastro, Johnston and Koribalski,
1996)  indicate a transverse velocity  at  least 20 times smaller than
required ($\ge$ 1\,000   km  s$^{-1}$).  PSR1706$-$44  has   been also
detected as a $\gamma$-ray  pulsar by the  EGRET instrument aboard the
Compton GRO (Thompson  et al. 1992)  and identified  with a the  COS-B
source 2CG342-02   (Swanenburg et al. 1981).   While  the other bright
EGRET  pulsars are   double-peaked  (see,  e.g., Thompson,    1998),the
$\gamma$ ray light curve of PSR1706$-$44  is characterized by a single
broad peak, offset in  phase relative to the  radio one.  PSR1706$-$44
has been  also detected  in  soft X-rays   by the  ROSAT/PSPC (Becker,
Brazier and Tr\"umper, 1995), as a weak non-pulsating source with a 18
\% upper limit on the  pulsed fraction. Pulsations were neither  found
in more recent ASCA, HRI (Finley  et al. 1998)  nor RXTE (Ray, Harding
and  Strickman, 1998) data.  The X-ray  spectrum of  the pulsar can be
described by a two  component model, i.e., a black  body at 0.4 keV in
the  ROSAT band plus  a power law (photon  index  $\sim$ 2.0) beyond 2
keV.   While the thermal component  may be  eventually due to emission
from the hot   neutron star surface,   the  non-thermal one  could  be
associated to a pulsar-powered synchrotron nebula, probably identified
in high  resolution ROSAT/HRI observations   (Finley et al.  1998). \\
From   a  general point   of view,   it   is interesting to note  that
PSR1706$-$44  has  many    characteristics in  common   with  the Vela
Pulsar. Their spin  periods  ($P \simeq  100$~ms)  as well as the  age
($\tau \sim 10^{4}$ yrs) and the rotational  energy loss ($\dot E \sim
10^{36}$  erg~s$^{-1}$)    are     similar.  Furthermore,        their
multiwavelength  behaviour  is  comparable, with a   similar spin-down
power conversion efficiency both in soft X-rays and high-energy ($\ge$
100 MeV) $\gamma$ rays. Assuming  that these similarities hold also in
the optical domain and scaling the  magnitude of Vela ($V=23.6$, d=500
pc) for the distance of PSR1706$-$44 (1.8  kpc, see Taylor \& Cordes,
1993), we get a rough magnitude of $V 
\sim 26$, to which a correction of at least one magnitude must be
added to  account for  the  higher interstellar absorption  (Finley et
al. 1998).  However,  we note  that this  estimate is  very  tentative
since the Vela  pulsar is still  the only case of  a $\sim 10^{4}$ yrs
neutron star detected  in the optical and  we  can not  exclude {\it a
priori} that PSR1706$-$44 be brighter. \\ In the optical, the field of
PSR1706$-$44 has been observed by our group both in 1993 and 1995 with
the ESO/NTT but no source was observed at the radio position (Johnston
et al. 1995)  down to a  limiting magnitudes of  $V \sim 24$ (Mignani,
1998) and $R \sim 24.5$  (unpublished).  The pulsar field was observed
again  by  Chakrabarty \& Kaspi  (1998)  who also  performed the first
optical timing experiment but no  pulsation was detected resulting  in
an upper limit of $R 
\ge 18$. \\
New observations of PSR1706$-$44 have been performed  last August as a
test case for  the Science Verification (SV) phase  of the First  Unit
(UT1) of  the ESO   Very  Large Telescope (Leibundgut,  De~Marchi  and
Renzini,  1998), aimed to assess  the scientific potentialities of the
telescope.   \\  The results  of  these observations  are  reported in
Sect. 2 and the results briefly discussed in Sect.3.


\section{Observations.}

The field of PSR1706$-$44 has been observed  on the night of Aug 17th,
1998 with the VLT-UT1 at the  ESO observatory in Paranal.  Observation
were performed in visitor mode by R.  Gilmozzi and B. Leibungut of the
SV team.   The UT1 was equipped with  a Test Camera,  commissioned for
the  SV program, mounting a  Tektronix   $2048^2$ CCD with a  measured
conversion  factor of   2.5 $e^{-}$/ADU and  a readout   noise  of 7.9
$e^{-}$ r.m.s.  The  CCD pixel size  is 27  $\mu$ which translates  to
0.0455 arcsec at the telescope plate scale, corresponding to a nominal
field        of   view   of     93     $\times$      93  arcsec   (see
http://www.eso.org/paranal/sv/ for  further details).  However, during
the science verification all observations were taken with a $2 
\times 2$ binning of the CCD, leading to an actual  pixel size of 0.09
arcsec.  A total of 6 exposures of 600 sec each has been obtained in a
Johnson V  filter.   Seeing conditions varied  from 0.5  to 0.6 arcsec
between single exposures, with an average airmass of 1.06. 

\begin{figure}
\centerline{\hbox{\psfig{figure=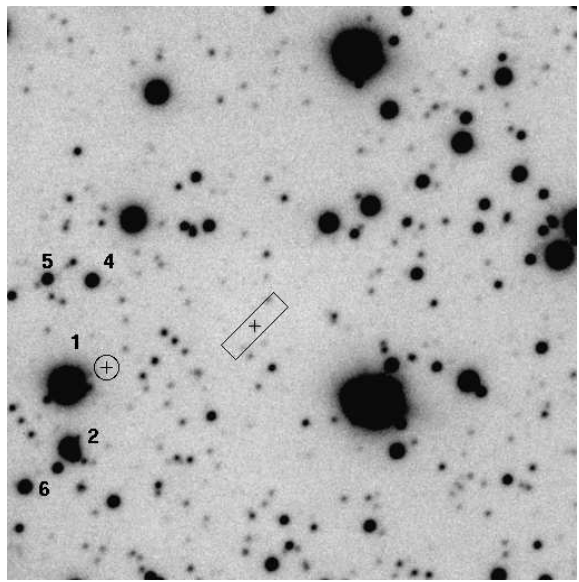,height=8cm,clip=}}}
\caption{V-band image of the field of PSR1706$-$44 taken with the
Test Camera of the VLT-UT1 for a total  exposure time of one hour. The
frame  is tilted  of  $\simeq 45\degr$ clockwise  wrt  the North. The
image size is $46 \times 46$ arcsec.  The two crosses mark, from right
to  left, the    pulsar  radio  positions according   to   Johnston et
al. (1995)-  position  \#1- and to  Frail \&  Goss (1998)/  Wang et al
(1988)- position  \#2. The   overall  uncertainty of our   astrometry,
resulting  both  from the errors on   the absolute coordinates  of our
reference stars  and on the r.m.s.  of  the  astrometric fit is $\simeq$
0.5 arcsec. The rectangle (1.6  $\times$ 6 arcsec in size) corresponds
to the error  region associated to position \#1  while the size of the
error circle around position   \#2  ($r \simeq   1.0$ arcsec)  has  been
confidently exaggerated. } 
\end{figure}

\subsection{Data reduction.}

Data reduction has been performed in Garching by members of the ESO SV
team. Particular care  was used for the basic reduction  steps since the CCD
suffered for color dependent blemishes,  with the largest one  located
practically at the  center.  Variations in  the CCD sensitivity caused
by dust grains were also noted by comparing flat field images taken in
different nights. All these problems made flatfielding very tricky and
requested a non-standard procedure.  Data  were thus flatfielded using
flats obtained    directly from the  sky  by  median-combining several
science exposures taken in  different nights.  The  use of  these {\it
superflats} lead  finally   to a flatfielding  accuracy  for wide-band
filters down to $1\%$, much higher than achievable either with dome or
twilight flats.  Bias frames were  obtained nearly every day and  show
no noticeable structure or   changes with time. Bias subtraction   and
flatfielding  were  performed     using   the  IRAF   {\it     ccdred}
package. Images were then corrected for the CCD dithering and combined
using the IRAF  tasks {\it imalign}  and {\it imcombine}.  Photometric
calibrations  have  been  performed    by imaging the   standard  star
PG1633+099 from a Landolt field.  The zero-point was computed applying
the IRAF  photometry package {\it digiphot},  with a final accuracy of
0.03 magnitudes. 

Astrometry on the image  has been performed using  as a  reference the
coordinates of  a few field  stars  extracted from  the USNO catalogue
using   the  ESO   Skycat     tool.  The   pixel-to-sky    coordinates
transformation has been thus  computed using the UK STARLINK  software
ASTROM (Wallace,  1990), leading to a  final accuracy  of $\simeq 0.5$
arcsec on our absolute astrometry  after taking in the consideration
both the r.m.s.  of the astrometric fit $(\simeq  0.3$ arcsec) and  the
average uncertainty on the USNO coordinates $(\simeq 0.25$ arcsec).

\subsection{Results.}

The most recent  radio position of PSR1706$-$44,  also reported in Tab.2
of the paper of Chakrabarty \& Kaspi (1998), has been obtained by
radio interferometry    measurements  (Frail  \& Goss,   1998)   and
independently confirmed  by pulsar  radio timing  (Wang et  al, 1998).
The  revised pulsar radio position has  thus  been superimposed on the
combined, one hour, V-band image of the field  (Fig.1). The star $(V =
17.4)$ close to the radio position is star 1 of Chackrabarty \& Kaspi
(1998). \\
We note that  the new  radio  coordinates of PSR1706$-$44 are   markedly
different ($\simeq 6$ and $\simeq 11$ arcsec  in RA and Dec, respectively)
from the old ones of Johnston et al. (1995), also superimposed in  Fig.1 for completeness. A blow-up
of the two radio error boxes is shown in Fig. 2a and b.
However, due to their much smaller error ($\sim  0.6$ arcsec) and to the higher
confidence on the  new measure,  we adopt the coordinates
of Frail \& Goss (1998)/Wang et al. (1998) -hereafter referred as \#2-
i.e.    $\alpha_{2000}$=17$^h$09$^m$42\fs73 and $\delta_{2000}$= -44\degr
29\arcmin07\farcs70.
Star F ($m_{V} = 25.9)$, visible at about two arcsec from the expected
pulsar position (Fig.  2b), is probably  too far to claim a convincing
association.  No other  object is visible  close to position \#2 down
to  a limiting magnitude of $V  \sim 27.5$. 

\begin{figure}
{\bf (a)}
\centerline{\hbox{\psfig{figure=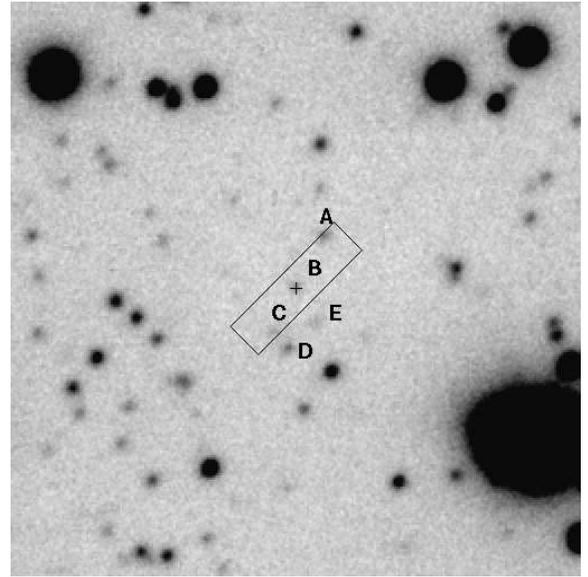,height=8cm,clip=}}}
{\bf (b)}
\centerline{\hbox{\psfig{figure=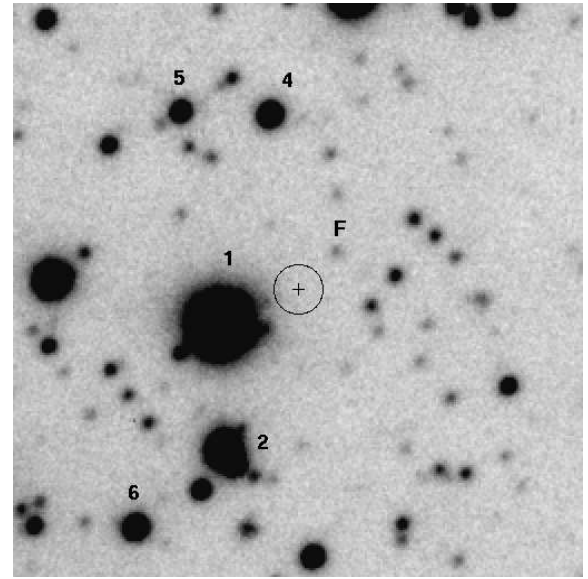,height=8cm,clip=}}}
\caption{Enlargement of Fig.1 around position \#1 (a) and position \#2
(b). Five objects with magnitudes between 25.7 and 26.8 are seen close
or inside the superseeded radio error box. No object is observed at
 position \#2 except for a $\sim 25.9$ mag object (star F) located
about 2 arcsec away.}
\end{figure}

\section{Conclusions}

We have presented the first deep search for a pulsar optical
counterpart performed with a 8 metre class telescope.  Although
these observations  with the UT1 of the VLT improve by more than two magnitudes our
previous imaging with the NTT, no optical counterpart to PSR1706$-$44 
has been detected down   to a limit   of  $V \sim 27.5$.     Only new
observations with the VLT-UT1, equipped with FORS1 could give higher  chances for
an optical detection of the pulsar. 

However, useful information  can be obtained also using the present  results.
According  to the Pacini law (Pacini,  1971),  the optical emission of
young pulsars is expected to be  purely magnetospheric and to decay on
a timescale  of few thousands years  at a  rate uniquely determined by
the  neutron star spin  parameters  and magnetic  field.  Although the
reality  of this,    so  called,  secular  decrease  has  never   been
convincingly proven (Nasuti    et     al. 1996)  it    always   seemed
circumstanciated by the case  of the Vela  Pulsar, which, being only a
factor 5 older than the Crab, is nearly 5  orders of magnitude weaker.
However, the lack of observational  data for other, similar,  $10^{4}$
yrs old neutron stars left the  evolution of the optical luminosity of
young  pulsars an  open point  for   a  long time, expecially in view
of the markedly different behaviour of middle-aged objects such
as PSR0656+14, Geminga and PSR1055-52 (Pavlov et al. 1997; Mignani et
al. 1998; Mignani et al. 1997). Our deep  optical
observations  of  PSR1706$-$44, comparable to  the  Vela pulsar in many
respects (age, timing, high-energy emission), provide now a new
piece of evidence to address this issue.  

In fact, in spite of the null result of these observations,
the magnitude upper limit, by far the lowest obtained so far from the ground
for an isolated neutron star, can be used to put contraints on the
optical luminosity of PSR1706$-$44 and thus to assess how this object
fits into the panorama of the optically emitting neutron stars.
Since the interstellar extinction expected from an hydrogen column density $10^{21}$ cm$^{-2}
\le N_{H} \le
5 ~10^{21}$ cm$^{-2}$  (Finley et al. 1998) spans in the
interval 0.6-3 magnitudes, the corresponding upper limit on the
optical luminosity of PSR1706$-$44 
can vary between $\simeq 2~10^{28}$ and $\simeq 5~10^{29}$ erg~s$^{-1}$. 
These values, although just upper limits, compare favourably with the
the general trend recognizable in Fig. 3, where we have plotted
the optical luminosity of the nine pulsars known to have
an optical counterpart (see e.g. Mignani, 1998) as a function of their
characteristic age. In particular, a turnover in the optical
luminosity seems to occur for pulsars aging around $10^{4}$ yrs. 
For older objects, the scenario
appears more  complicated by  the onset  of  thermal emission from the
neutron  star  surface  which,  as in   the   case of the  middle-aged
PSR0656+14 (Pavlov et al. 1997) and Geminga (Mignani et al. 1998), can
significantly,   if not  complitely, account for   the overall optical
luminosity. 

\begin{figure}
\centerline{\hbox{\psfig{figure=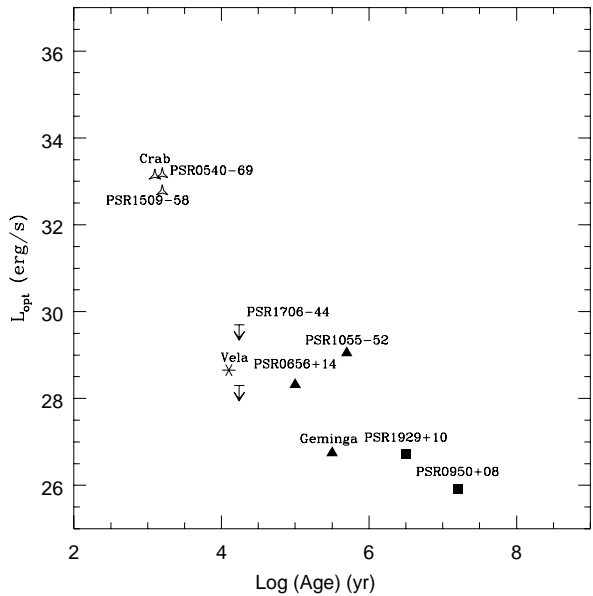,height=8cm,clip=}}}
\caption{The upper limit range for the optical luminosity of
PSR1706$-$44 is plotted as a function of the spin-down age 
together with the actual values for the pulsars with an associated (proposed) optical
counterpart (see Mignani, 1998 and references therein). 
Young ($\sim 10^{3}$ yrs), middle-aged ($\sim 10^{5}$ yrs) and old
($\ge 10^{6}$
yrs) objects are indicated by open triangles,full triangles,  and filled squares, respectively.
Vela is marked by an asterisk. }
\end{figure}

\begin{acknowledgements}
R. Mignani is  grateful to the SV  team for the  details given both on
the observation and on the  data reduction. A  special thank is due to
Dick Manchester for confirming the pulsar radio  position and to Nichi
D'Amico for   useful suggestions. \\  We  also aknowledge  the support
software  provided by the  Starlink Project which  is funded by the UK
SERC. 

\end{acknowledgements}

\end{document}